\documentclass[sigconf,10pt]{acmart}
\makeatletter\def\@authorfont{\Large}\makeatother

\usepackage{xcolor}
\usepackage{graphicx}
\usepackage{booktabs}
\usepackage{enumitem}
\usepackage{xspace}
\usepackage{array}
\usepackage{pifont}
\usepackage{tikz}
\usepackage{pgfplots}
\pgfplotsset{compat=1.17}
\usepackage{graphicx}
\usetikzlibrary{positioning, arrows.meta, shapes.geometric, fit, calc, backgrounds}

\usepackage{microtype}
\usepackage{newfloat}
\DeclareFloatingEnvironment[name=Listing,placement=t]{listing}
\definecolor{ptblue}{HTML}{4477AA}
\definecolor{ptcyan}{HTML}{66CCEE}
\definecolor{ptgreen}{HTML}{228833}
\definecolor{ptyellow}{HTML}{CCBB44}
\definecolor{ptred}{HTML}{EE6677}
\definecolor{ptpurple}{HTML}{AA3377}
\definecolor{ptgrey}{HTML}{BBBBBB}

\newcommand{\sys}{\textsc{Pramana}\xspace}
\newcommand{\smartparagraph}[1]{\par\noindent\textbf{#1}\ }

\newcommand{\yes}{\textcolor{ptgreen}{\ding{51}}}
\newcommand{\no}{\textcolor{ptred!80!black}{\ding{55}}}
\newcommand{\pmk}{\textcolor{ptyellow!80!black}{\raisebox{-1pt}{$\sim$}}}

\usepackage{xcolor}

\newcommand{\NumPapers}{66}
\newcommand{\NumES}{387}

\newcommand{\NumInScope}{255}
\newcommand{\PctInScope}{66}

\newcommand{\NumImplSat}{87}
\newcommand{\PctImplSat}{34}
\newcommand{\NumImplGap}{168}

\renewcommand{\NumImplSat}{68}
\renewcommand{\PctImplSat}{27}
\renewcommand{\NumImplGap}{168}
\newcommand{\NumImplUndetermined}{19}

\newcommand{\PctNetreplicaSat}{11}

\newcommand{\PctMininetSat}{9}

\newcommand{\NumMultiAxis}{192}
\newcommand{\PctMultiAxis}{75}
\newcommand{\NumGapInfra}{81}
\newcommand{\NumGapBuild}{82}

\newcommand{\NumGapMPTCP}{23}
\newcommand{\NumGapPhones}{18}
\newcommand{\NumGapCellular}{13}
\newcommand{\NumLinksNeed}{38}
\newcommand{\NumLinksParallel}{30}
\newcommand{\NumLinksSeries}{8}
\newcommand{\NumLinksMininet}{5}
\newcommand{\NumLinksNeedBuild}{26}
\newcommand{\NumLinksNeedTrace}{6}

\newcommand{\NumGapMultiMissing}{84}
\newcommand{\NumGapAdditions}{81}
\newcommand{\NumGapMultiUse}{56}

\newcommand{\NumGapBBRtwo}{20}

\newcommand{\NumTopologyOnly}{12}

\newcommand{\PctCoverGreedyHead}{53}

\title{\sys: A Composable, Domain-Specific Backend for Empirical Networking Research}
\author{\mbox{Jaber Daneshamooz}}
\affiliation{\institution{UC Santa Barbara}\country{}}

\author{\mbox{Eugene Vuong}}
\affiliation{\institution{UC Santa Barbara}\country{}}

\author{\mbox{Alagappan Ramanathan}}
\affiliation{\institution{UC Irvine}\country{}}

\author{\mbox{Manni Moghimi}}
\affiliation{\institution{UC Santa Barbara}\country{}}

\author{\mbox{Haarika Manda}}
\affiliation{\institution{UC Santa Barbara}\country{}}

\author{\mbox{Satyam Kumar}}
\affiliation{\institution{IIT Delhi}\country{}}

\author{\mbox{Snithik Thode}}
\affiliation{\institution{IIT Delhi}\country{}}

\author{\mbox{Satyandra Guthula}}
\affiliation{\institution{UC Santa Barbara}\country{}}

\author{\mbox{Sylee Beltiukov}}
\affiliation{\institution{UC Santa Barbara}\country{}}

\author{\mbox{Dongsu Han}}
\affiliation{\institution{KAIST}\country{}}

\author{\mbox{Tarun Mangla}}
\affiliation{\institution{IIT Delhi}\country{}}

\author{\mbox{Sangeetha Abdu Jyothi}}
\affiliation{\institution{UC Irvine}\country{}}

\author{\mbox{Walter Willinger}}
\affiliation{\institution{Northwestern University}\country{}}

\author{\mbox{Arpit Gupta}}
\affiliation{\institution{UC Santa Barbara}\country{}}

\copyrightyear{2026}
\acmYear{2026}
\setcopyright{cc}
\setcctype{by}
\acmConference[HotNets '26]{The 25th ACM Workshop on Hot Topics in Networks}{November 16--17, 2026}{Salt Lake City, UT, USA}
\acmBooktitle{The 25th ACM Workshop on Hot Topics in Networks (HotNets '26), November 16--17, 2026, Salt Lake City, UT, USA}
\acmDOI{10.1145/3847290.3859953}
\acmISBN{979-8-4007-3029-0/2026/11}
\begin{document}
\begin{abstract}
Empirical networking research often tests a hypothesis with data the study itself generates, so accelerating this research requires shortening the time from forming a hypothesis to generating that data.
Consider whether a bulk BBR download fairly shares its bottleneck with a Google Meet call.
Testing this requires configuring a realistic bottleneck link, running the BBR transfer and the Meet call together, and collecting the call's quality metrics.
Today a researcher builds this setup from scratch for each new idea; this delay from idea to data is the \emph{ideation-to-data gap}.
AI agents already propose hypotheses faster than researchers can generate data to test them. Closing the gap needs a backend that generates the data for each new hypothesis on demand, without a new setup.

We envision \sys, a composable \emph{backend} for networking research designed as a thin waist, with research intents above it and execution substrates below.
At its core, the \emph{intent specification} separates the data a study needs (the \emph{intent}) from where that data is generated (the \emph{substrate}) and how it is produced (the \emph{mechanism}), so one specification can target any substrate.
We build a corpus of \NumInScope\ data-generation intents from \NumPapers\ published papers and find that the intent specification expresses all of them, while no existing tool covers more than \PctNetreplicaSat\%.
Our proof-of-concept implementation covers \PctImplSat\% of these intents. For CCAnalyzer, one of the \NumPapers\ papers, the implementation regenerated the data for one trial of each network setting and congestion-control algorithm in the paper's main experiment.
We lay out what implementing the full design requires and why it needs a community effort.

\end{abstract}

\renewcommand{\shortauthors}{Daneshamooz et al.}
\maketitle

\section{Introduction}
\label{sec:intro}



Empirical networking research tests hypotheses against data, so its speed depends on how long it takes to go from an idea to data that can confirm or refute it.
AI agents now propose hypotheses far faster than anyone can test them~\cite{glia,aiscientist,aiscientist_v1,airesearcher,alphaevolve}, and testing them has become the bottleneck~\cite{popper}.




\begin{figure}[t]
  \centering
\includegraphics[width=\linewidth]{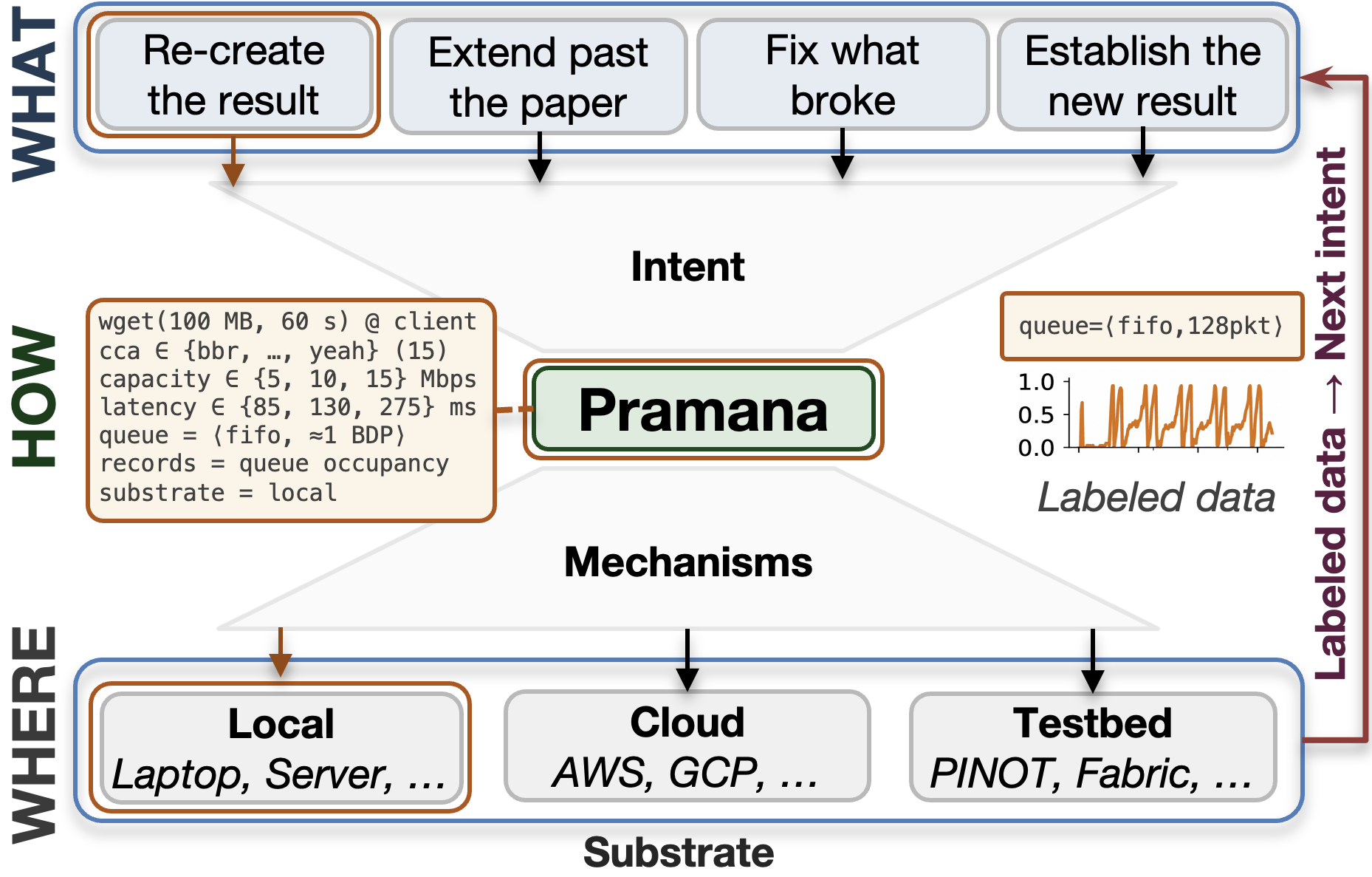}
  \caption{\textbf{Data-generation thin waist.}}
  \label{fig:arch}
  \vspace{-2em}
\end{figure}
An empirical study draws on four recurring activities~\cite{lakatos}: re-creating a published result, extending it until it breaks, fixing what broke, and establishing a new one.
Each activity needs data, and a researcher requests that data by issuing an \emph{intent}.
In networking, an intent specifies what data to generate, while a \emph{substrate} specifies where to generate it (e.g., a laptop, the cloud, or a testbed). An intent names the bottleneck's capacity and queue, the cross-traffic pressure on it, and the \emph{workflows}, the programs that run on its end hosts. A workflow either runs a command-line tool, such as an \texttt{iperf3} bulk download, or is an \emph{application workflow} that drives a real application through a session, such as Google~Meet. Workflows send traffic through the bottleneck, record what the study measures, or do both. A packet capture is a \emph{telemetry workflow}, and so is the Meet call when it reports its video statistics. Each workflow that sends traffic carries the delay and loss of its own path and the congestion control of its sender.
Every intent fills these same fields with different values.
Stating an intent takes a few sentences, but realizing it means writing code against several tools.
A few groups amortize that cost with verticals built for one domain, such as Pantheon~\cite{pantheon} for congestion control. Intents span too many domains to build one vertical for each, so most researchers rebuild the setup per study.
We call the resulting delay between an idea and its data the \emph{ideation-to-data gap}.

Over the past fifteen years, related efforts have narrowed this gap. Mininet~\cite{mininet,mininethifi}, netUnicorn~\cite{netunicorn}, NetReplica~\cite{netreplica,daneshamooz2026netreplicalong}, and NetGent~\cite{netgent} each turned one concern (topology, node mapping, bottleneck conditions, and application workflow) into a reusable, replayable artifact.
Yet each tool by itself is insufficient: NetReplica cannot drive an application, and NetGent cannot shape a bottleneck.
Running a real Google~Meet call against a BBR bulk download over one shared, shaped bottleneck, for instance, takes several of these tools at once.
The gap persists because these tools make a researcher write the intent and the substrate into the \emph{mechanism}, the code that produces the data.
Composing them still forces a researcher to re-encode the intent in one platform's mechanism, so the intent is tied to that platform and must be rewritten for the next one.
A thin waist (Figure~\ref{fig:arch}) would place one specification between research intents and execution substrates and compose these capabilities behind it.



\sys (Sanskrit for \emph{evidence}) designs this thin waist around the \emph{intent specification}.
\sys{} applies \emph{progressive disaggregation}~\cite{netreplica}, separating coupled parts one step at a time so that each separated part can then be specified and changed on its own.
The Internet's designers took such a step when they separated TCP from IP, so that applications TCP did not suit, such as real-time speech, could use IP without TCP's reliable delivery~\cite{clark88}.
\sys{} composes NetReplica's bottleneck conditions, in which cross-traffic recorded on production networks replays through an emulated bottleneck, with NetGent's deterministically replayed application workflows (e.g., YouTube, Zoom, and Twitch).
It also borrows netUnicorn's separation of what an experiment runs from where it runs.
\sys{} adds one more separation: it moves both the what and the where into the specification, out of the mechanism.
A researcher can then change the bottleneck, a workflow, or the substrate without changing the other two.
Because each of the four activities changes some of these parts and keeps the rest, one specification is designed to serve all four activities on any substrate.
A researcher writes the intent in natural language, and a model compiles it into the specification. Only that step uses a language model.
Everything below it composes and replays existing artifacts and logs each run's requested conditions.

This paper makes four contributions:

\begin{itemize}[leftmargin=1.2em,itemsep=0pt,topsep=2pt]
  \item \textbf{The intent corpus.} We build the first corpus of research intents, \NumInScope\ intents from \NumPapers\ measurement papers published over a decade. It grounds the intent specification in real demand, can be reused and extended by others, and is the yardstick against which we measure the design, today's implementation, and each specialized tool (\S\ref{sec:eval}).
  \item \textbf{The abstraction.} We introduce the \emph{intent specification}, a domain-specific language in which a researcher states the intent (\emph{what}) and the substrate (\emph{where}); the backend supplies the mechanism (\emph{how}). Separating the \emph{what} and the \emph{where} from the \emph{how} is a step of progressive disaggregation that NetReplica, NetGent, and netUnicorn do not take. The language expresses every in-scope intent in the corpus. Because every field maps to a capability of an existing system, we also use it to score what each tool and today's implementation can generate (\S\ref{sec:pramana}, \S\ref{sec:eval}).
  \item \textbf{The architecture.} We design an architecture that composes NetReplica and NetGent behind versioned capability files, in which each service declares what it supports, and that, like netUnicorn, keeps a workflow separate from the node it runs on (\S\ref{sec:pramana}).
  \item \textbf{A demonstration on published intents.} We test the backend on the corpus intents instead of ones we designed. We measure coverage across the corpus against each specialized tool and regenerate the data of CCAnalyzer's~\cite{ccanalyzer} main experiment, one trial per pairing of setting and algorithm.
\end{itemize}

\noindent
The intent corpus and the related reproducibility artifact are publicly available~\cite{pramana-artifact}.

\section{Background and Motivation}
\label{sec:motivation}


\smartparagraph{The four activities.}
Each study re-creates, extends, fixes, or establishes a result~\cite{lakatos}, and a recent pair of papers from our corpus (\S\ref{sec:eval}) does all four:
CCAnalyzer~\cite{ccanalyzer} identifies a flow's congestion control algorithm by how it fills a bottleneck queue, reaching 100\% accuracy across 15 algorithms;
its successor, CClinguist~\cite{cclinguist}, \emph{re-creates} that classifier as a baseline.
It \emph{extends} the classifier until it breaks, as CCAnalyzer's accuracy drops to 84.3\% at a latency outside its fixed profiles.
It \emph{fixes} the break by generating those profiles automatically, recovering accuracy to 91.7\%, and it \emph{establishes} new labels on algorithms outside CCAnalyzer's set, including the first ECN-based ones.
Each activity needs its own data, and CClinguist's new labels are what a later study would re-create.

\smartparagraph{Intent, substrate, mechanism.}
These activities need different data, but each needs controlled data on demand, with every sample labeled by the conditions that produced it (its ground truth).
Each request names an intent and a substrate and relies on a \emph{mechanism} (\S\ref{sec:intro}).
The intent and substrate are easy to state, but the mechanism means standing up the bottleneck, driving the real application, imposing the cross-traffic, and wiring the telemetry, then repeating it all for the next substrate.

\smartparagraph{The thin waist.}
An intent is built from bottleneck conditions and workflows, telemetry included, and a node mapping places its workflows on a substrate's nodes. These parts recur across studies and substrates, so a shared backend that builds each once and reuses it for any specification on any substrate shrinks the ideation-to-data gap.
Such a backend is the \emph{thin waist}~\cite{netunicorn} of Figure~\ref{fig:arch}, a narrow layer between many intents and many substrates.

\smartparagraph{The requirements.}
Table~\ref{tab:requirements} lists the seven requirements this backend must meet and scores each system against them.
The first, \emph{flexibility}, has three dimensions, and the other six (\emph{portability}, \emph{usability}, \emph{fidelity}, \emph{evolvability}, \emph{robustness}, and \emph{scale}) make the backend practical for researchers to run.
Each listed predecessor meets at most one dimension of flexibility and exposes a tool-specific interface, and only our composed design meets all seven.

\begin{table}[t]
  \centering
  \footnotesize
  \setlength{\tabcolsep}{4pt}
  \begin{tabular}{@{}>{\raggedright\arraybackslash}p{0.42\columnwidth}ccccc@{}}
    \toprule
    \textbf{Requirement} & \textbf{Mnt} & \textbf{nUn} & \textbf{nRp} & \textbf{nGt} & \textbf{\sys} \\
    \midrule
    \multicolumn{6}{@{}l@{}}{\textit{Flexibility} --- the wide top} \\
    \quad Diverse network env & \pmk & \no  & \yes & \no  & \yes \\
    \quad Complex app workflow   & \no  & \no  & \no  & \yes & \yes \\
    \quad Node mapping                & \no  & \yes & \no  & \no  & \yes \\
    \midrule
    Portability across substrates     & \no  & \yes & \no  & \no  & \yes \\
    Usability of the intent interface & \pmk & \pmk & \pmk & \pmk & \yes \\
    Fidelity of spec to intent        & \no  & \no  & \no  & \no  & \yes \\
    Evolvability of subsystems        & \no  & \no  & \no  & \no  & \yes \\
    Robustness under failure          & \no  & \no  & \no  & \no  & \yes \\
    Scale to research-time sweeps     & \no  & \no  & \no  & \no  & \yes \\
    \bottomrule
  \end{tabular}
\caption{\textbf{Requirements for a data-generation thin waist and the systems that meet them.}
  \yes:~native \quad \pmk:~partial \quad \no:~absent. \quad Mnt:~Mininet, nUn:~netUnicorn, nRp:~NetReplica, nGt:~NetGent. \sys's column scores the design; \S\ref{sec:pramana} states what the implementation has. 
  }
  \label{tab:requirements}
  \vspace{-3em}
\end{table}

\smartparagraph{Progressive disaggregation.}
Each tool in Table~\ref{tab:requirements} applies progressive disaggregation~\cite{netreplica} to one part of an experiment.
Mininet~\cite{mininet,mininethifi} separates the topology from the hardware, and netUnicorn~\cite{netunicorn} separates an experiment's tasks from the nodes that run them.
NetReplica~\cite{netreplica} separates the bottleneck's static settings from its dynamic cross-traffic, and that cross-traffic from the network that recorded it.
NetGent~\cite{netgent} breaks an application workflow into reusable states that compose into deterministic replays.
\sys's design separates the intent and its substrate from the mechanism: each field names a capability that a service declares, and the backend picks that service, so a specification names no mechanism.

\smartparagraph{A motivating example.}
Prudentia's~\cite{prudentia} intent, also in our corpus (\S\ref{sec:eval}),
simultaneously needs the application, the bottleneck, and the mapping of its workflows to nodes.
The \emph{Google~Meet} call (WebRTC, up to 1.5~Mbps) plays \emph{``the reference Big Buck Bunny video''} and contends with the BBR download on one shared drop-tail bottleneck of \emph{``approximately 4$\times$BDP''} at 8 or 50~Mbps with a normalized 50~ms RTT, and every sample carries \emph{``video resolution, frames per second, freezes per minute, and high-delay packets''} (those past the ITU 190~ms bar)~\cite{prudentia}.
NetReplica shapes the bottleneck, replays cross-traffic, and runs client commands, but it cannot join a call or any other interactive session, so the call's quality never appears.
NetGent drives the call but has no contended bottleneck to put it behind, and Mininet runs a real client but cannot replay recorded cross-traffic.
The researcher therefore builds a per-paper orchestration script that puts both workflows behind one shared bottleneck. NetReplica sets the drop-tail bottleneck, NetGent runs the call, and the script sets BBR on the download's sender through \texttt{sysctl}, as our CCAnalyzer script did (\S\ref{sec:eval}).
The script also sets each workflow's delay, starts the download, the capture, and the call under study,
and labels each sample with its conditions.
The intent leaves settings implicit: the 4$\times$BDP queue is 200~KB at 8~Mbps and 1.25~MB at 50~Mbps, and a 50~ms RTT means adding whatever delay the substrate's path lacks.
Writing the script requires learning each tool's interface and parameters, and every later study repeats that work.
To move the study to AWS or replace the call with a Twitch stream, the researcher rewrites the script.
We use Prudentia's intent as the running example of what the design must express, and CCAnalyzer's intent as the one whose data we regenerate (\S\ref{sec:eval}).

\smartparagraph{What the abstraction adds.}
\sys{} is designed to replace this described 
script with one specification.
In its grammar (Listing~\ref{lst:grammar}), the Meet call is the Foreground, the workflow under study, and the BBR download is a second workflow that the Regime, the imposed conditions, loads onto the same bottleneck as Dynamic pressure, with a separate path and congestion control for each workflow.
An orchestration script over the same tools reveals a missing capability (e.g.,  BBR for a remote sender) only in the data it returns.
The design checks a specification against the declared capabilities before anything runs.
With each field tied to a capability, one grammar scores every tool against the same 255 intents from 66 papers (\S\ref{sec:eval}).

\section{Pramana: A Data-Generation Thin Waist}
\label{sec:pramana}


We first describe the design; the last two paragraphs state what the implementation runs today.

\smartparagraph{The intent specification.}
\sys's abstraction, the \emph{intent specification}, composes the artifacts that the tools of \S\ref{sec:motivation} disaggregated.
An experiment is then a \emph{composition} of existing artifacts, a $\mathit{Foreground}$ of workflows running over a $\mathit{Regime}$ of imposed conditions, and the researcher only states the intent.
We write the abstraction as a domain-specific language whose grammar Listing~\ref{lst:grammar} gives.
A model translates the researcher's natural-language intent, written by hand or extracted from a paper, into this language.

\begin{listing}[t]
  \centering
  \footnotesize
  \setlength{\tabcolsep}{2pt}
  \begin{tabular}{@{}r@{~}c@{~}l@{}}
    $\mathit{Evidence}$ & $=$ & $\mathit{collect}(\mathit{ExperimentSet})$ \\
    $\mathit{ExperimentSet}$ & $=$ & $\mathit{compile}(\mathit{Intent},\mathit{Answers})~|~\text{user-written}~\mathit{Spec}$ \\
    $\mathit{Experiment}$ & $=$ & $\langle \mathit{Foreground} \otimes \mathit{Regime} \rangle \times \mathit{iterations}$ \\
    $\mathit{Foreground}$ & $=$ & $\{~\mathit{Workflow}_i~@~\mathit{node}_i~\}$ \\
    $\mathit{Workflow}$ & $=$ & $\mathit{CLI}(\mathit{app})~\triangleright~\mathit{NFA}~~\text{with}~\mathit{path},\mathit{cca}$ \\
    $\mathit{path}$ & $=$ & $\langle \mathit{latency}, \mathit{jitter}, \mathit{loss}, \mathit{reorder}, \mathit{dup} \rangle$ \quad (per workflow) \\
    $\mathit{cca}$ & $=$ & $\langle \mathit{algo}, \mathit{stack}, \mathit{version}, \mathit{params}, \mathit{flags}, \mathit{build} \rangle$ \\
    $\mathit{Regime}$ & $=$ & $\mathit{Static}\langle \mathit{capacity}{\downarrow\uparrow}, \mathit{queue} \rangle \otimes \mathit{Dynamic}\langle \mathit{pressure} \rangle$ \\
    $\mathit{queue}$ & $=$ & $\langle \mathit{discipline}, \mathit{size}, \mathit{params}, \mathit{ecn}, \mathit{mode} \rangle$ \\
    $\mathit{pressure}$ & $=$ & $\mathit{replay}(\mathit{ctp})~|~\mathit{load}(\mathit{Workflow}^{*} \mapsto \mathit{Node}^{*})~|~\text{both}$ \\
    $\mathit{iterations}$ & $=$ & $\mathit{int}~|~\mathit{until}(\mathit{sufficient}(\mathit{signal},\mathit{precision}),\mathit{min},\mathit{max})$ \\
    $\mathit{identity}(e)$ & $=$ & $H(\mathit{workflow@sha},\mathit{params},\mathit{path},\mathit{cca},\mathit{static},\mathit{dynamic})$ \\
    $\mathit{collect}$ & $=$ & $\mathit{deploy};\mathit{prepare};\mathit{verify};\mathit{run};\mathit{publish}$ \\
  \end{tabular}
  \caption{\textbf{The intent-specification grammar.}}
  \vspace{-2em}
  \label{lst:grammar}
\end{listing}

\smartparagraph{The intent-specification grammar.}
Listing~\ref{lst:grammar} defines how evidence is produced: $\mathit{compile}()$ produces a specification from the intent, the middle rows define its structure, and $\mathit{collect}$, defined in the last row, is the mechanism that realizes it.
The bottleneck ($\mathit{Regime}$) is precise and shared, and NetReplica imposes it once for all workflows. Each $\mathit{path}$ is approximate and belongs to one workflow, as does its sender's congestion control ($\mathit{cca}$).
An $\mathit{NFA}$ is a nondeterministic finite automaton over an application session, and a $\mathit{ctp}$ is a recorded cross-traffic trace that NetReplica replays.
The $\mathit{until}$ clause stops a run once a signal the application or tool reports is statistically sufficient, so no stopping rule depends on a computed result.

A specification denotes a distribution over condition-labeled datasets, each sample tagged with the condition vector it realized.
These precise semantics let the backend decide before any run whether the declared services can realize a specification (its \emph{realizability}). Because a specification can name only what some service has declared, realizability reduces to a satisfiability check over those declarations that runs in polynomial time, whereas for open-ended code the same question is undecidable~\cite{capisce,minesweeper}.
The semantics also keep everything below \emph{compile} deterministic, since $\mathit{compile}()$ is the only step that may invoke a model. Determinism covers how an experiment is built and driven. Live traffic still varies, as a live call differs between runs, and each sample's label records the conditions that run actually realized.
The intent specification stops at data generation, because pulling metric extraction into it would fix a set of metrics, narrow the intents a researcher can express, and require the model to reason past \emph{compile}, breaking the determinism below it.
\sys{} produces the data for the confirmed intent, and the researcher decides whether it supports the hypothesis.

\smartparagraph{Compiling an intent.}
For the Prudentia intent of \S\ref{sec:motivation}, \sys{} compiles the researcher's description clause by clause into the grammar of Listing~\ref{lst:grammar}.
The Meet call, a browser session that plays the reference video, becomes the $\mathit{Foreground}$, a $\mathit{Workflow}$ pinned to a node.
The shaped bottleneck becomes the $\mathit{Regime}$, whose $\mathit{Static}$ part sets the intent's two capacities and the $\mathit{queue}$ (a drop-tail FIFO of $\approx 4\times$BDP).
$\mathit{Dynamic}$ $\mathit{pressure}$ loads a second $\mathit{Workflow}$, a bulk \texttt{iperf3} download whose sender runs $\mathit{cca}=\mathrm{BBR}$, onto the bottleneck.
Each workflow's $\mathit{path}$ carries the normalized 50~ms latency.

\begin{figure}
    \centering
    \includegraphics[width=\linewidth]{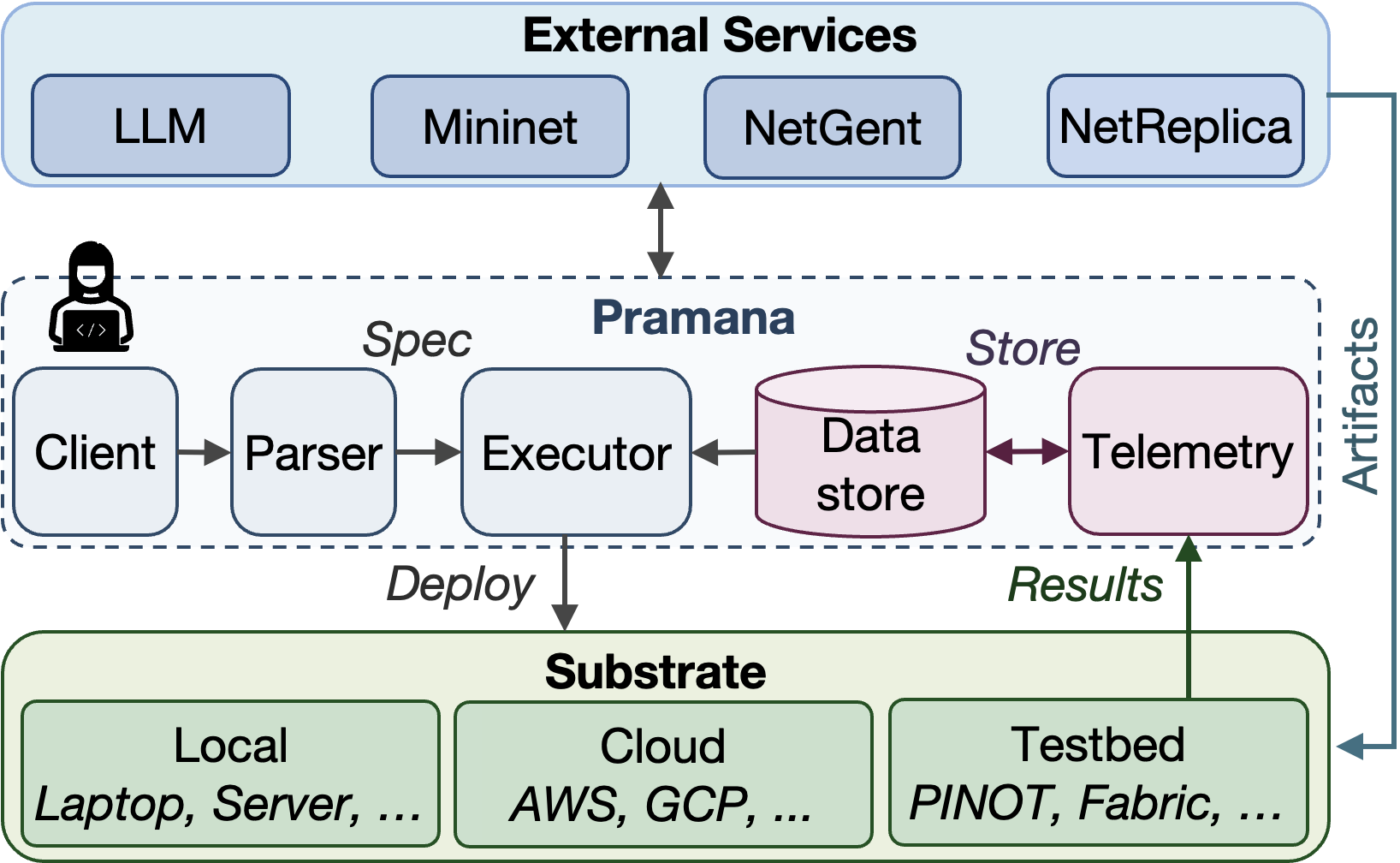}
\caption{\textbf{\sys's architecture.}}
\label{fig:architecture}
\vspace{-0.6em}
\end{figure}

\smartparagraph{The architecture.}
In the design (Figure~\ref{fig:architecture}), the model compiles the intent into a specification, and the Parser checks every field against the capabilities NetGent and NetReplica declare, confirms the translation with the researcher, and gives each experiment a content-hash identity.
The Executor is designed to realize each experiment on any substrate: local (a laptop or desktop), cloud (AWS~\cite{aws}, GCP~\cite{gcp}), or a testbed (PINOT~\cite{pinot}, RIPE~Atlas~\cite{ripeatlas}, FABRIC~\cite{fabric}, CAIDA~Ark~\cite{caidaark}). NetGent and NetReplica ship their artifacts straight to the substrate, and the nodes report results to Telemetry, which stores them.

\smartparagraph{The design choices.}
In the design, Prudentia's specification keeps the Meet workflow, the shaped bottleneck, and the node mapping as separate artifacts. A researcher can therefore swap the call for a Twitch stream without touching the bottleneck (\emph{flexibility}) and move the run from a laptop to AWS by changing only the node mapping (\emph{portability}).
A researcher expresses intents in natural language; the model only fills in a fixed grammar, so its output can be checked field by field (\emph{usability}).
Every service lists its capabilities in a versioned capability file, and the parser rejects any field that names a capability no file lists. It then echoes the translation in plain language for the researcher to confirm, so the researcher can spot mistranslated values and workflows that the check cannot see (\emph{fidelity}).
Because these files are versioned, NetGent can add a workflow, such as one that records a new metric, without changing the grammar (\emph{evolvability}).
A specification names every input, so each experiment's identity is a hash of those inputs (Listing~\ref{lst:grammar}).
The backend is crash-only: after any crash, it restarts and reruns only the experiments with no stored result (\emph{robustness}).
A failed run keeps its data and a failure label.
The Executor fences off a worker that loses contact with it, so that worker cannot disturb another experiment's conditions.
When an application's interface changes and a NetGent workflow no longer matches it, NetGent regenerates the workflow offline, and once the researcher confirms it, the workflow runs under a new hash.
The hash also lets the backend skip identical experiments, and a persistent worker pool amortizes setup across a sweep (\emph{scale}).

\smartparagraph{The implementation today.}
Our proof of concept connects to NetReplica and NetGent through their service interfaces, runs in local containers and on AWS, and regenerated CCAnalyzer's traces through a short script that set each run's parameters (\S\ref{sec:eval}).
It offers a single shaped bottleneck for each run's traffic, 22 NetGent application workflows, and 15 kernel congestion-control algorithms. It shapes no second link, though \NumLinksNeed\ in-scope intents need several (\S\ref{sec:eval}).
For CCAnalyzer's runs, the implementation records the requested conditions and the queue trace, with its own bandwidth and latency probes off during capture so that their traffic does not enter the trace.
The implementation has none of the design's robustness and scale mechanisms, so each experiment waits 44~s for a fresh worker, a crashed campaign cannot resume, and identical experiments run again. Under bursts of concurrent workers, the single telemetry service limits capture-heavy studies.

\smartparagraph{Running the Prudentia intent.}
For Prudentia's intent, the compile step fills each unstated field with a default and reports that interpretation back with the results.
NetReplica shapes the drop-tail bottleneck on the substrate worker.
NetGent, however, runs one workflow per run, either from its library or, for a shell tool such as \texttt{iperf3}, generated on request. So the Meet call and the \texttt{iperf3} download have not run together on the link, and the implementation reports none of the call metrics that \S\ref{sec:motivation} lists.
BBR can be set two ways, through the kernel and per socket. The implementation applies both, and only for a sender inside the substrate.
It rejects an algorithm that is not included in the 15 offered ones or that the kernel cannot load, and it fails when asked for an unknown queue discipline.
The only combination of fields it checks is a queue setting that contradicts the buffer size, so a request for BBR on a remote sender runs without BBR and without a warning.
The design's check needs a capability file from each service, and no service publishes one today.

\section{Evaluation}
\label{sec:eval}


\begin{figure}[t]
\centering
\includegraphics[width=0.9\columnwidth]{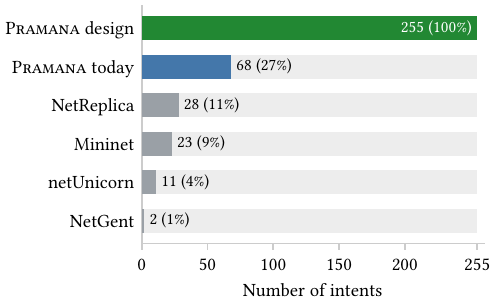}
\caption{\textbf{In-scope intents that each system covers.}}
\label{fig:coverage}
\vspace{-0.6em}
\end{figure}

\begin{listing}[t]
\centering
\footnotesize
\fbox{\parbox{\columnwidth}{\raggedright
Through a FIFO bottleneck sized near one BDP, at 5, 10, or 15~Mbps and 85, 130, or 275~ms of RTT, download a 100~MB file with \texttt{wget} for 60~s from a server running each of the 15 Linux congestion-control algorithms, five samples each, and record when packets enter, leave, and drop at the bottleneck.}}

\vspace{1pt}
\setlength{\tabcolsep}{2pt}
\begin{tabular}{@{}r@{~}c@{~}l@{}}
$\mathit{Experiment}$ & $=$ & $\langle \mathit{Foreground} \otimes \mathit{Regime} \rangle \times 5$ \\
$\mathit{Foreground}$ & $=$ & $\{\,\texttt{wget}(100\,\mathrm{MB}, 60\,\mathrm{s})\ @\ \mathit{client},\ \texttt{http\_server}\ @\ \mathit{server}$ \\
 & & \quad with $\mathit{cca} \in \{$bbr, bic, cdg, cubic, \ldots, yeah$\}$ (15)$\,\}$ \\
$\mathit{Regime}$ & $=$ & $\mathit{Static}\langle \mathit{capacity} \in \{5, 10, 15\}\,\mathrm{Mbps},\ \mathit{queue} = \langle \mathrm{fifo}, \approx 1\,\mathrm{BDP} \rangle \rangle$ \\
$\mathit{path}$ & $=$ & $\langle \mathit{latency} \in \{85, 130, 275\}\,\mathrm{ms~RTT} \rangle$ \\
\textit{records} & $=$ & queue occupancy at the bottleneck \\
\end{tabular}
\vspace{-1em}
\caption{\textbf{An intent and its specification.}
Top: the intent we extract for the runs that test CCAnalyzer's classifier in its main experiment (1 of its 11). Bottom: the specification it compiles to (Listing~\ref{lst:grammar}): 9 network settings $\times$ 15 algorithms $\times$ 5 samples $=$ 675 runs.}
\label{lst:example}
\vspace{-1.55em}
\end{listing}

We ask three questions: (i)~whether the grammar expresses every intent we collected, and how its coverage compares with existing specialized tools, (ii)~how much of the corpus the implementation covers today, and (iii)~what regenerating one published dataset costs.

\smartparagraph{Experimental setup.}
Our unit of analysis is the \emph{experiment}, the data collection behind one measured result.
We sample papers by CCS concept and keyword, such as \emph{Network measurement}, \emph{Transport protocols}, and \emph{congestion control} across SIGCOMM, IMC, and NSDI over the past decade.
We build the corpus by hand from these papers, identifying each paper's results and, for each result, the experiments that produced it, and record them in a Result$\times$Experiment matrix.
We write down each experiment's intent as natural-language text, the way the paper states it.
An intent is in scope when its experiment generates its own traffic, which we call \emph{endogenous} data generation, excluding simulation, passive capture, and offline computation.
Of \NumES\ extracted intents, \NumInScope\ (\PctInScope\%) are in scope and form the in-scope corpus, one intent per experiment.
CCAnalyzer~\cite{ccanalyzer}, for example, reports 26 data-bearing figures and tables, and they come from 11 experiments. Its headline result (100\% classification accuracy) rests on one of them, 675 runs over nine network settings and 15 congestion-control algorithms.
Prudentia~\cite{prudentia} reports its findings from 8 experiments, one of which pits a Meet call against competing downloads, the intent of \S\ref{sec:motivation}.
Listing~\ref{lst:example} shows CCAnalyzer's intent as we extract it and its specification.
A grammar axis is a group of related fields, such as the bottleneck or the application workflow, and most intents (\NumMultiAxis, \PctMultiAxis\%) need two or more.
For each in-scope intent, one reader records its needs from the paper and a second checks every entry against the same text; a deterministic rule then compares them with a capability sheet that cites each system's code.

\smartparagraph{Counting rule.}
We score every system the same way, \sys's implementation included. A system covers an intent when its released code, as of the HotNets submission, can create everything the intent needs through the system's own interface, without code written for that intent.
Standard command-line tools, such as \texttt{iperf3}, count for every system that runs commands on a host. Software that must be compiled for the intent, including a kernel other than the stock one, does not count.
Driving an application through its interface is a NetGent feature, so it counts for NetGent and for \sys{} through NetGent.
If the paper emulates an intent's conditions, a system that recreates them covers it wherever the paper ran it, and replaying recorded cross-traffic through a fixed-rate bottleneck recreates a time-varying capacity. If it measures a real network or runs on phones, the system must reach that network or those devices.
If the paper leaves a need unstated, such as the algorithm its server ran, we cannot decide the intent for any system and report it separately.
Counting an intent does not mean we have run it, and this section generates data only for CCAnalyzer.

\smartparagraph{Coverage of existing tools.}
Figure~\ref{fig:coverage} scores each specialized tool against the corpus, and every miss traces to a grammar axis the tool lacks. NetReplica imposes bottleneck conditions but does not drive an application workflow. netUnicorn places an experiment but does not impose its network conditions, and Mininet executes the application in an isolated namespace but cannot drive it as an application workflow or impose trace-based cross-traffic~\cite{netreplica,netburst}.
The best tool, NetReplica, covers \PctNetreplicaSat\% of the corpus and Mininet \PctMininetSat\%, and the tools miss different axes, each lacking one that another supplies.
The grammar is the union of those axes and expresses all \NumInScope\ in-scope intents from \NumPapers\ papers.

\begin{figure}[t]
\centering
\includegraphics[width=.9\linewidth]{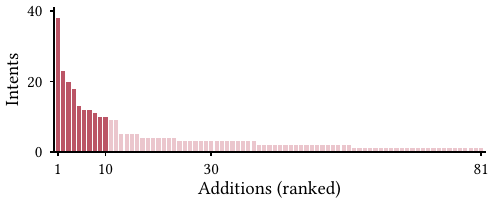}
\caption{\textbf{Intents that need each missing addition.}}
\label{fig:needs}
\vspace{-1.2em}
\end{figure}

\smartparagraph{Implementation coverage.}
\sys's implementation covers \NumImplSat\ of the \NumInScope\ intents today (\PctImplSat\%). Another \NumImplUndetermined\ are underspecified, so we cannot decide them.
Of the \NumImplGap\ intents it cannot cover, \NumGapBuild\ need a congestion-control or transport build and \NumGapInfra\ need infrastructure it cannot reach (an intent can need both), and half (\NumGapMultiMissing) lack more than one addition.
They need \NumGapAdditions\ additions in all, and \NumGapMultiUse\ of them serve more than one intent (Figure~\ref{fig:needs}).
The largest is several shaped links for one intent, each with its own capacity, queue, and loss: \NumLinksParallel\ need two parallel paths for a multipath transport, and \NumLinksSeries\ need links in series, such as an access link and an interconnect.
Mininet builds such topologies but covers \NumLinksMininet\ of them: of the rest, \NumLinksNeedBuild\ also lack a software build, such as an MPTCP kernel or a QUIC stack, and \NumLinksNeedTrace\ a recorded capacity trace on a path.
For \NumTopologyOnly\ intents, the shaped links are the only missing addition.
Next come two kernels compiled from publicly available code, an out-of-tree MPTCP kernel (\NumGapMPTCP) and a BBRv2 kernel (\NumGapBBRtwo), and two connectors that \sys{} needs to reach phones (\NumGapPhones) and cellular access (\NumGapCellular).
Added in the order that unblocks the most intents, ten additions raise coverage from \PctImplSat\% to \PctCoverGreedyHead\%.

\smartparagraph{Regenerating CCAnalyzer's data.}
We pick one corpus paper, CCAnalyzer, and \sys{} regenerates the queue-occupancy traces its classifier reads in its main experiment, with one trial for each setting and each of the 15 algorithms.
In a separate 20-run instrumented benchmark, the compile takes 2.7~s and about 10,000 tokens on a small language model (Gemini 3.1 Flash-Lite). This 2.7~s is the intent-to-specification compile time with artifacts and infrastructure already in place, and it excludes the rest of the researcher's effort.
In a user study, 46 students used \sys{} to regenerate this result, and 89\% reported that they would have needed at least twice the effort to match it by hand~\cite{pramana-edu}.
Each run's 60~s measurement and everything after it run unattended. Getting the data \emph{in hand} adds a 71~s telemetry transfer, which streaming the upload would shorten.
In our runs, the script supplied the link parameters, the algorithm, and the download workflow as structured fields that overrode the compile's output. The download's server ran on the host machine, outside the substrate worker, and the driver set that server's algorithm through \texttt{sysctl}.
A CCAnalyzer author judged the regenerated traces similar to the originals. The sender's segmentation offload also queued packets in bundles of two or more segments, so each queue held at least double its configured number of packets.

\section{Related Work}
\label{sec:related}


\smartparagraph{Emulators, testbeds, and verticals.}
Link and testbed emulators shape a bottleneck's bandwidth, delay, loss, and queue: Dummynet~\cite{dummynet} and netem~\cite{netem} on one link, Emulab~\cite{emulab}, ModelNet~\cite{modelnet}, and DieCast~\cite{diecast} across a topology, Mininet~\cite{mininet,mininethifi} in software, and Mahimahi~\cite{mahimahi} by record-and-replay.
Each shapes the link and leaves the real application and the cross-traffic to the researcher.
Verticals such as Puffer~\cite{puffer} and Pantheon~\cite{pantheon} fix one application and metric set (adaptive bitrate or congestion control) and accelerated research in that domain.
\sys{} builds on their fidelity and expresses intents from many domains in one grammar.

\smartparagraph{Agentic accelerators and language interfaces.}
Agentic research systems stack hypothesis generation on top, a domain backend in the middle, and raw instruments below.
AI~Scientist~\cite{aiscientist} works at the top layer, and systems research suits its propose-and-test loop because a proposed solution can be checked by measuring it~\cite{barbarians}.
In networking, NetArena scores automation agents against network emulators~\cite{netarena}. None of these systems composes the instruments into a backend that turns an intent into controlled, ground-truth-labeled data.
\sys's compile step draws on semantic parsing and text-to-code generation~\cite{spider,codex}. Grammar-constrained decoding rules out ill-formed output by construction~\cite{picard,gcd,outlines}, and a human who confirms the translation catches the errors it cannot~\cite{translation-validation,necula}.
In intent-based networking, the closest work, Lumi, reads its translation of a configuration back to the operator for confirmation~\cite{lumi}, and \sys{} applies the same check before generating a condition-labeled dataset.

\section{Discussion and Conclusion}
\label{sec:discussion}


\smartparagraph{Checking published evidence.}
\sys{} can help artifact evaluation, whose committees stop at \emph{Functional} or \emph{Reusable} badges. These badges confirm that a paper's code runs but not that its results hold, because re-collecting the data is too costly~\cite{saucez-ccr2019,willinger-credibility-ccr2025}.
Regenerating that data from an intent is the \emph{re-create} activity (\S\ref{sec:motivation}) applied to another paper.
\sys{} supplies the data and makes every unstated field an explicit default, possibly unlike the paper's original condition. Whether the data supports the paper's finding remains an analysis question.
Airavat~\cite{airavat} generates Internet-measurement analysis pipelines and verifies them against established methodology. Paired with \sys's regenerated data, it would take a paper's intent to a checked result.

\smartparagraph{Artifacts.}
\sys{} generates data only from artifacts that already exist.
Creating them from a paper's description currently lies outside its scope.
That being said, debugging a dysfunctional yet publicly accessible artifact entails mechanical tasks of debugging an existing artifact, setting up its software environment, and validating that it runs.
LLM agents have been shown to perform reasonably well for these tasks~\cite{cvegenie}.
We envision NetGent could be extended to take up this task, but it does not support it right now.
Though we have not measured the cost, we expect this cost to be one-time per artifact because a finished workflow or build serves every intent that needs it.

\smartparagraph{Where does AI belong?}
As models improve, an agent that writes and runs each experiment end to end grows more tempting than \sys's single, bounded compile step.
Does bounding a model with a fixed grammar use AI well, or discard the generality that makes agents useful?


\smartparagraph{What should shared infrastructure admit?}
One intent or agent can launch thousands of experiments, yet public infrastructure assumes a researcher who reserves nodes, files requests, and schedules runs.
What load limits should RIPE~Atlas~\cite{ripeatlas}, FABRIC~\cite{fabric}, and campus testbeds declare, so agents can check bandwidth commitments, reservation caps, and contention before launching, without human approval?

\smartparagraph{Conclusion.}
\sys{} is a thin waist for empirical networking research: a researcher states which data to collect and on which substrate, and \sys{} supplies the mechanism.
Its intent specification can express all \NumInScope\ corpus intents, while NetReplica, the best existing tool, covers \PctNetreplicaSat\%.
A language model compiles a plain-language description into the specification, and the backend runs it with NetGent and NetReplica.
Today, the implementation covers \NumImplSat\ intents and has regenerated CCAnalyzer's main experiment once per setting and algorithm.
Of the remaining intents, 168 require missing capabilities and 19 are underspecified. Building the ten most critical ones would lift coverage from \PctImplSat\% to \PctCoverGreedyHead\%, but long-term sustainability will only derive from community building.
With these capabilities, a researcher or an agent gets the data for a hypothesis by writing its intent instead of building a setup.
That ability shortens the ideation-to-data gap for both research~\cite{daneshamooz2026rethinking} and teaching~\cite{pramana-edu}.

\smartparagraph{Acknowledgments.}
We thank our shepherd, Soudeh Ghorbani, the anonymous reviewers, and the CCAnalyzer authors for their feedback. Researchers at UCSB have been supported by NSF awards 2443777 and 2323229 and by research gifts from Cisco and Google. Researchers at UCI have been supported by NSF award 2443024.
\newpage
\bibliographystyle{ACM-Reference-Format}
\bibliography{hotnets25-template}

\end{document}